\pdfoutput=1
\documentclass[12pt]{article}

\font\grande=cmr9.5 scaled \magstep4
\font\medio=cmr9.5 scaled \magstep2
\outer\def\beginsection#1\par{\medbreak\bigskip
      \message{#1}\leftline{\bf#1}\nobreak\medskip
\vskip-\parskip
      \noindent}
\usepackage{graphicx} 
\begin{document}
\bibliographystyle {unsrt}

\titlepage
\begin{flushright}
CERN-TH-2018-106
\end{flushright}

\vspace*{1.5cm}
\begin{center}
{\grande Blue and violet graviton spectra}\\
\vspace{5mm}
{\grande from a dynamical refractive index}\\
\vspace{15mm}
 Massimo Giovannini 
 \footnote{Electronic address: massimo.giovannini@cern.ch} \\
\vspace{1cm}
{{\sl Department of Physics, CERN, 1211 Geneva 23, Switzerland }}\\
\vspace{0.5cm}
{{\sl INFN, Section of Milan-Bicocca, 20126 Milan, Italy}}

\vspace*{1cm}
\end{center}

\centerline{\medio  Abstract}
\vspace{5mm}
We show that the spectral energy distribution of relic gravitons mildly increases for frequencies 
smaller than the $\mu$Hz and then flattens out whenever the refractive index of the tensor modes 
is dynamical during a quasi-de Sitter stage of expansion. For a conventional thermal history 
the high-frequency plateau ranges between the mHz and the audio band 
but it is supplemented by a spike in the GHz region if a stiff post-inflationary phase precedes 
the standard radiation-dominated epoch.  Even though  the slope is
blue at intermediate frequencies,  it may become violet in the MHz window.  
For a variety of post-inflationary histories, including the conventional one, a dynamical index of refraction
leads to a potentially detectable spectral energy density in the kHz and in the mHz regions while 
all the relevant phenomenological constraints are concurrently satisfied. 

\vskip 0.5cm

\nonumber
\noindent

\vspace{5mm}

\vfill
\newpage
Relic gravitons are copiously produced in the early Universe because of the pumping action of the 
background geometry \cite{zero}. If a conventional stage of inflationary expansion is suddenly 
replaced by a radiation-dominated epoch, the spectral energy density in critical units at the present 
conformal time $\tau_{0}$ (denoted hereunder by $\Omega_{gw}(\nu,\tau_{0})$) 
is quasi-flat \cite{one} for comoving frequencies\footnote{The scale factor shall be normalized throughout 
as $a(\tau_{0}) =a_{0} =1$. Hence, within the present notations, comoving and physical frequencies do coincide 
at the present time.} $\nu$ ranging, approximately, between $100$ aHz and $100$ MHz. 
The transition across the epoch of matter-radiation equality leads to an infrared branch where 
$\Omega_{gw}(\nu,\tau_{0}) \propto \nu^{-2}$ between the aHz and $100$ aHz \cite{one}. 
If the post-inflationary plasma is stiffer than radiation (i.e. characterized by a barotropic index 
$w = p/\rho$ larger than $1/3$) the corresponding spectral energy density inherits a blue 
(or even violet) slope for typical frequencies larger than the mHz and smaller 
than about $100$ GHz \cite{two}. 

Gravitational waves might however acquire an effective index of refraction when they travel in curved 
space-times \cite{three} and their spectral energy distribution becomes comparatively larger than in the 
conventional situation \cite{four}. If the refractive index
increases during a quasi-de Sitter stage of expansion, the propagating speed 
diminishes and $\Omega_{gw}(\nu,\tau_{0}) \propto 
\nu^{n_{T}}$ (with $n_{T}>0$) for $\nu$ ranging between $100$ aHz and $100$ MHz 
(recall that $1\, \mathrm{aHz} = 10^{-18} \mathrm{Hz}$). There 
 are no compelling reasons why there should be a single increasing branch extending 
 throughout the whole  range of variation of the comoving frequency.
On the contrary we shall show that there are regions in the parameter space where all the phenomenological 
constraints are concurrently satisfied while the spectral energy distribution is only blue in the intermediate frequency range
(roughly speaking below the $\mu$Hz) while it flattens out (and it may even decrease) around the mHz 
band. Depending on the post-inflationary thermal history the high-frequency 
plateau may reach out deep into the audio band and beyond.
\begin{figure}[!ht]
\centering
\includegraphics[height=5.7cm]{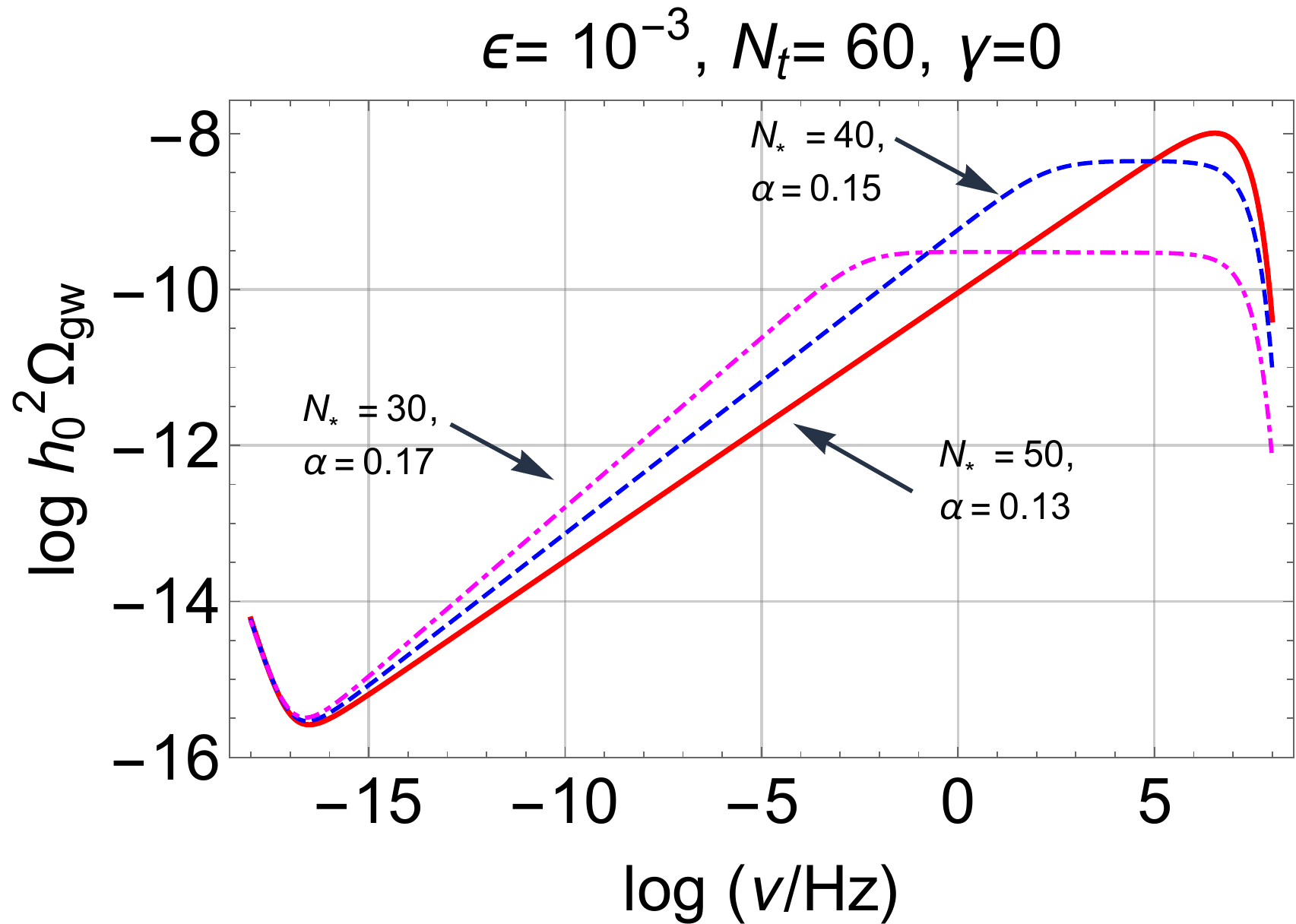}
\includegraphics[height=5.7cm]{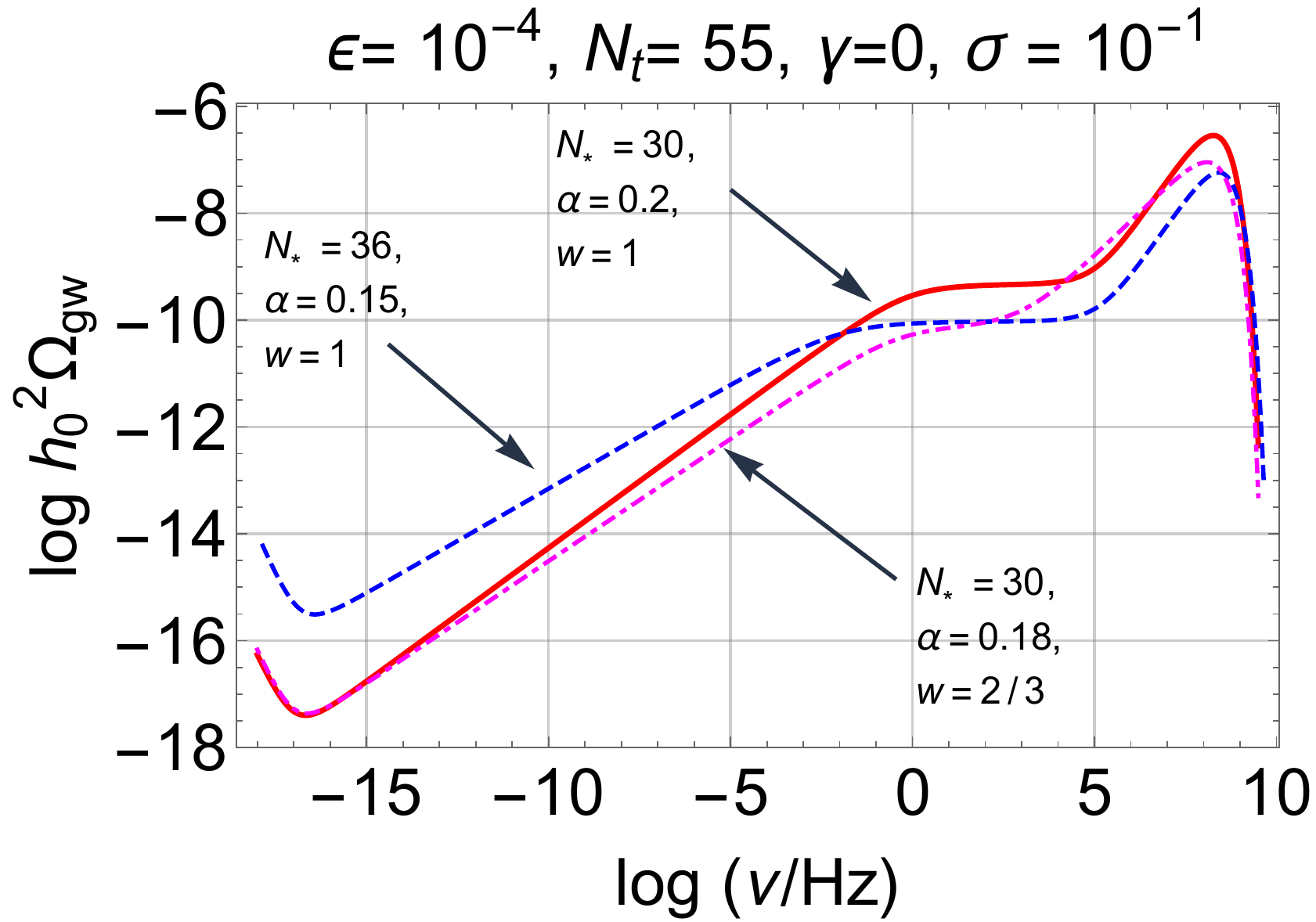}
\caption[a]{The spectral energy distribution of the relic gravitons produced 
by the variation of the refractive index during inflation is illustrated as a function of the comoving 
frequency for two broad classes of post-inflationary evolutions. The 
energy density is measured in critical units and common logarithms are employed on both axes. }
\label{FIGU1}      
\end{figure}

The spectral energy distribution of the relic gravitons produced by the 
variation of the refractive index is illustrated in Fig. \ref{FIGU1} where, in both plots, $N_{t}$ and 
$N_{*}= a_{*}/a_{i}\leq N_{t}$ denote, respectively, the total and the critical number of efolds beyond 
which the refractive index goes back to $1$. Note also that $\epsilon$ is the slow-roll parameter 
while $\gamma$ is a parameter appearing in the modified action for the tensor modes in the 
presence of a dynamical refractive index (see below Eq. (\ref{EQ3}) and discussion therein).
While the explicit evolution may vary \cite{three,four}, the results of Fig. \ref{FIGU1} refer to the situation where 
the refractive index evolves as a power of the scale factor $a$ during inflation\footnote{The rate of variation of the refractive index during the inflationary stage of expansion is given by $\alpha$ in units of the Hubble rate;
note also that, by definition, $n_{*} = n_{i} (a_{*}/a_{i})^{\alpha}$ with $n_{i} = 1$. } i.e. $n(a) = n_{*} (a/a_{*})^{\alpha}$ for $a< a_{*}$, 
while  $n(a) \to 1$ when $a > a_{*}$.  An explicit profile with this 
property is $n(x) = (n_{*} x^{\alpha} e^{- \xi x} +1)$ where $x = a/a_{*}$; $\xi$ controls 
the sharpness of the transition and we shall bound the attention to the case $\xi>1$ 
(in practice $\xi  = 2$ even if larger values of $\xi$ do not change the conclusions reported here). 

According to Fig. \ref{FIGU1}, if $N_{*}$ is just slightly smaller than $N_{t}$ 
the spectral energy density is increasing in the whole range 
of comoving frequencies between\footnote{Note that $k_{eq} = 0.0732\,[h_{0}^2 \Omega_{R0}/(4.15\times 10^{-5})]^{-1/2}\,\,  h_{0}^2 \Omega_{M0} \,\, \mathrm{Mpc}^{-1}$ where $\Omega_{M0}$ and $\Omega_{R0}$ are the values of the critical fractions of matter and radiation in the concordance paradigm;  $h_{0}$ denotes the present value of the Hubble rate $H_{0}$ in units of $100 \,\mathrm{km}/(\mathrm{sec}\,\times\mathrm{Mpc})$. If $h_{0}^2 \Omega_{M0} = 0.1411$ \cite{fivea,five},  $\nu_{eq} = k_{eq}/(2\pi) = {\mathcal O}(100) \, \mathrm{aHz}$.} $\nu_{eq}= {\mathcal O}(100) \, \mathrm{aHz}$ and $\nu_{\mathrm{max}}= {\mathcal O}(200)\, \mathrm{MHz}$. As  soon as $N_{*}$ diminishes substantially, $\Omega_{gw}(\nu,\tau_{0})$
develops a quasi-flat plateau whose slope is controlled by the slow-roll parameter $\epsilon$ (see dashed and dot-dashed curves in the left plot of Fig. \ref{FIGU1}).  The knee and the end point of the spectral energy distribution are fixed by the typical frequencies $\nu_{*}$ and $\nu_{\mathrm{max}}$:
\begin{eqnarray}
\nu_{*} &=& p(\alpha, \epsilon, N_{*}, N_{t}) \, \nu_{\mathrm{max}}, \qquad  
p(\alpha, \epsilon, N_{*}, N_{t})= \biggl| 1 + \frac{\alpha}{1- \epsilon}\biggr| e^{N_{*}(\alpha+ 1) - N_{t}},
\label{ONEa}\\
\nu_{\mathrm{max}} &=& 1.95\times 10^{8} \biggl(\frac{\epsilon}{0.001}\biggr)^{1/4} 
\biggl(\frac{{\mathcal A}_{{\mathcal R}}}{2.41\times 10^{-9}}\biggr)^{1/4} 
\biggl(\frac{h_{0}^2 \Omega_{R0}}{4.15 \times 10^{-5}}\biggr)^{1/4}  \,\, \mathrm{Hz},
\label{ONEb}
\end{eqnarray}
where ${\mathcal A}_{{\mathcal R}}$ denotes the amplitude 
of the power spectrum of curvature inhomogeneities at the wavenumber $k_{p} =0.002\,\mathrm{Mpc}^{-1}$ \cite{fivea,five}
corresponding to the pivot frequency $\nu_{p} = k_{p}/(2 \pi) =  3.092\,\,\mathrm{aHz}$ that defines the infrared 
band of the spectrum. In the right plot of Fig. \ref{FIGU1} the spike at the end of the quasi-flat plateau
is characterized by the frequency $\nu_{spike}= \nu_{\mathrm{max}}/\sigma > \nu_{\mathrm{max}}$ (with $\sigma<1$)
extending into the GHz band.  This occurrence reminds of the conventional situation when the refractive index is not dynamical: 
in this case the value of the end-point frequency of the spectrum depends on the post-inflationary 
thermal history \cite{two} and may exceed $\nu_{\mathrm{max}} = {\mathcal O}(200)$ MHz.
All in all Fig. \ref{FIGU1} demonstrates that the spectral energy density may consist of two, three or even 
four different branches: the infrared branch (between the aHz and $100$ aHz) is supplemented by a second mildly 
increasing branch extending between $10^{-16}$ Hz and $200$ MHz. 
Furthermore, when the variation of the refractive index terminates before the end of inflation 
a third branch develops between $\nu_{*}$ and $\nu_{\mathrm{max}}$.
Finally if the post-inflationary evolution is dominated, for some time, 
by a stiff barotropic fluid, then a fourth branch arises between few kHz 
and the GHz region.

The results summarized by Fig. \ref{FIGU1} follow from the action describing the evolution of the tensor modes of the 
geometry in the presence of a dynamical refractive index \cite{three,four}:
\begin{equation}
S=  \frac{1}{8 \ell_{P}^2} \int d^{3} x \int d\tau \,\,a^2\,n^{2\gamma}\biggl[ \partial_{\tau} h_{ij} \partial_{\tau} h_{ij} 
- \frac{\partial_{k} h_{ij} \partial^{k} h_{ij}}{n^2} \biggr],\qquad \ell_{P} = \sqrt{8 \pi G} = \frac{1}{\overline{M}_{P}},
\label{EQ3}
\end{equation}
where $a(\tau)$ denotes the scale factor of a conformally flat geometry of Friedmann-Robertson-Walker type\footnote{More specifically  $\overline{g}_{\mu\nu}= a^2(\tau) \eta_{\mu\nu}$ and
$\eta_{\mu\nu}=\mathrm{diag}(1\,, -1\, -1, -1)$ is the Minkowski metric.} and, as already mentioned, $n(\tau)$ is the index of refraction.
The parameter $\gamma$ accounts for different possible parametrization of the 
effect: for instance, motivated by the original suggestion of Ref. \cite{three},  the first paper of Ref. \cite{four} 
suggested an action (\ref{EQ3}) with $\gamma = 0$; after this analysis some other authors (see e.g. second and third papers 
of Ref. \cite{four}) considered a model with $\gamma = 1$. These two cases are just related by a conformal rescaling. Moreover
for a generic case $\gamma \neq 0$ the slope of $\Omega_{gw}(\nu,\tau_{0})$ is the one obtainable for  $\gamma =0$  up to a $\gamma$-dependent
 prefactor that can be tacitly absorbed in a redefinition of the spectral index. 
 Thus, even if the pivotal case is $\gamma =0$ Eq. (\ref{EQ3}) encompasses all the
different possibilities and accounts for their physical equivalence.

For immediate convenience the conformal time coordinate $\tau$ can be traded for the newly defined $\eta$-time whose 
explicit definition is $n(\eta) d\eta = d\tau$;  in the $\eta$-parametrization Eq. (\ref{EQ3}) becomes:
\begin{equation}
S=  \frac{1}{8 \ell_{P}^2} \int d^{3} x \int d\eta \,\,b^2(\eta) \,\,\biggl[\partial_{\eta} h_{ij} \partial_{\eta} h_{ij} - \partial_{k} h_{ij} \partial^{k} h_{ij} \biggr],
\qquad b(\eta) = a\, n^{\gamma -1/2}.
\label{EQ4}
\end{equation}
Note that Eqs. (\ref{EQ3})--(\ref{EQ4}) reproduce the results of Ref. \cite{foura} in the limit $n \to 1$. 
Furthermore, since $n(a)$ 
goes back to $1$ before the end of inflation, $\eta$ and $\tau$ eventually 
coincide in the post-inflationary phase so that the system can be directly quantized and 
solved in the $\eta$-time where the mode expansion for the field operator reads:
\begin{equation}
\hat{h}_{ij}(\vec{x}, \eta) = \frac{\sqrt{2} \ell_{P}}{(2\pi)^{3/2} b(\eta)}\sum_{\lambda} \int \, d^{3} k \,\,e^{(\lambda)}_{ij}(\vec{k})\, \biggl[ f_{k,\lambda}(\eta) \hat{a}_{\vec{k}\,\lambda } e^{- i \vec{k} \cdot \vec{x}} + f^{*}_{k,\lambda}(\eta) \hat{a}_{\vec{k}\,\lambda }^{\dagger} e^{ i \vec{k} \cdot \vec{x}}\biggr].
\label{EQ5}
\end{equation}
In Eq. (\ref{EQ5}) $\lambda= \oplus,\, \otimes$ runs over the tensor polarizations but, as in the conventional situation, the evolution of the mode functions is the same for each of the two values of $\lambda$: 
\begin{equation}
\ddot{f}_{k} + \biggl[ k^2 - \frac{\ddot{b}}{b} \biggr] f_{k} =0,\qquad \dot{f} = \frac{\partial f}{\partial \eta} \equiv \frac{1}{n} \,\frac{\partial f}{\partial \tau} = \frac{f^{\prime}}{n}.
\label{EQ6}
\end{equation}
The overdot denotes here a derivation with respect to $\eta$ ({\em not} with respect to the cosmic time coordinate, as often tacitly assumed) 
while the prime denotes, as usual, a derivation with respect to the conformal time coordinate $\tau$. Simple algebra shows that
$\ddot{b}/b= {\mathcal F}^2 + \dot{{\mathcal F}}$; note that ${\mathcal F}= \dot{b}/b$ 
coincides with ${\mathcal H}= a^{\prime}/a$ in the limit $n \to 1$ by virtue of the basic relation $n(\eta)  d\eta = d\tau$
that connects Eqs. (\ref{EQ3}) and (\ref{EQ4}).

Equation (\ref{EQ6}) is equivalent to an integral equation with initial conditions assigned at $\eta_{ex}$, where $\eta_{ex}$ is the turning point defined by the condition $k^2 = \ddot{b}_{ex}/b_{ex}$: 
\begin{eqnarray}
f_{k}(\eta) &=& \frac{b}{b_{ex}} \biggl\{ f_{k}(\eta_{ex}) 
+ \biggl[ \dot{f}_{k}(\eta_{ex}) - {\mathcal F}_{ex} f_{k}(\eta_{ex})\biggr] \int_{\eta_{ex}}^{\eta} \frac{b_{ex}^2}{b^2(\eta_{1})} d\eta_{1}
\nonumber\\
&-&k^2\, b_{ex}\, \int_{\eta_{ex}}^{\eta} \frac{d\eta_{1}}{b^2(\eta_{1})} \int_{\eta_{ex}}^{\eta_{1}} b(\eta_{2}) f_{k}(\eta_{2}) d\eta_{2} \biggr\}.
\label{INT1}
\end{eqnarray}
When $\eta  < \eta_{*}$ (i.e. $N_{t} < N_{*}$) we have that $b(\eta) = b_{*} (- \eta/\eta_{*})^{-\delta}$ 
where $b_{*} = a_{*} n_{*}^{\gamma -1/2}$ and $\delta = [2 + \alpha (2 \gamma -1)]/[2 ( 1 + \alpha - \epsilon)]$; in this regime $\ddot{b} \neq 0$
so that the turning point is $k \eta_{ex} = {\mathcal O}(1)$ and $\eta_{ex} \simeq 1/k$.  Similarly, 
when $a > a_{*}$ we have $\ddot{b}/b= a^{\prime\prime}/a \simeq (2 - \epsilon)/[\tau^2(1 - \epsilon)^2]$ 
so that, again, $k \tau_{ex} ={\mathcal O}(1)$. If the reentry takes place when $\ddot{b} \neq 0$ the relevant turning points are determined 
by the condition $k^2 =|\ddot{b}_{re}/b_{re}|$, i.e. $k \eta_{re} = {\mathcal O}(1)$. However, if the 
reentry occurs during a radiation-dominated stage of expansion, Eq. (\ref{EQ6}) implies instead 
$\ddot{b} = a^{\prime\prime} \to 0$: since the curvature coupling vanishes 
in the vicinity of the second turning point $\tau_{re}$ the condition 
$k^2 = \ddot{b}_{re}/b_{re} \to 0$ implies $k \eta_{re} \ll 1$ (and not, as it could be naively guessed, 
$k \eta_{re} = {\mathcal O}(1)$).  

From  Eq. (\ref{INT1}) the spectral energy distribution can be analytically estimated
 by matching the lowest-order solution across $\eta_{re}$ and by evaluating the obtained result when the 
corresponding wavelengths are all shorter than the Hubble radius
at the present epoch:
\begin{eqnarray}
\Omega_{gw}(k, \tau_{0}) &=& \frac{k^4}{12 H^2 \overline{M}_{P}^2 a^4 \pi^2}
\biggl[1 + \biggl( \frac{{\mathcal F}_{ex}}{k} \biggr)^2 \biggr]\biggl(\frac{b_{re}}{b_{ex}}\biggr)^2 \biggl[1 + b_{ex}^4{\mathcal J}^2(\eta_{ex}, \eta_{re})\biggr],
\label{INT3a}\\
{\mathcal J}(\eta_{ex}, \eta_{re}) &=& \int_{\eta_{ex}}^{\eta_{re}} \frac{d \eta_{1}}{b^2(\eta_{1})}.
\label{INT3b}
\end{eqnarray}
Whenever  $\eta_{ex} < \eta_{*}$ (and the reentry takes place during radiation), Eqs. (\ref{INT3a})--(\ref{INT3b}) imply that  
$\Omega_{gw} \propto \nu^{\overline{n}_{T}}$ where\footnote{Note that since $\epsilon < \alpha <1$ 
the spectral index is given, to leading order, by $\overline{n}_{T}  \simeq (3 - 2\gamma) \alpha$: as anticipated after Eq. (\ref{EQ3})
different values of $\gamma$ simply rescale the value of $\alpha$.} $ \overline{n}_{T} = 2 ( 1 - \delta) 
\equiv [\alpha ( 3 - 2 \gamma) - 2 \epsilon]/(1 + \alpha - \epsilon)$: 
this is the slope appearing in both plots of Fig. \ref{FIGU1} for $\nu< \nu_{*}$.
Conversely, if $\eta_{ex}  >\eta_{*}$ (and the reentry takes place during radiation) the spectral energy density 
scales as $\Omega_{gw} \propto \nu^{- 2\epsilon}$: this is the quasi-flat slope illustrated in both plots of Fig. \ref{FIGU1}
for $ \nu >\nu_{*}$. Finally the MHz branch depends on the 
post-inflationary thermal history which modifies the spectrum whenever $\eta_{re}$ does not 
fall within the radiation epoch: indeed the presence of a stiff 
phase preceding the radiation stage introduces a further branch 
corresponding to the modes reentering after the end of inflation and 
before the radiation dominance. In this branch spectral index is  
$\overline{n}_{T} = [4 - 2/(1-\epsilon) - 4/(3 w +1)]$; if, for instance, $w \to 1$ 
the spectral index becomes explicitly $\overline{n}_{T} \to 1 + {\mathcal O}(\epsilon)$: this is, incidentally, the slope
 of $\Omega_{gw}(\nu,\tau_{0})$ before the spike in the GHz band 
 (see, in this respect, the right plot of Fig. \ref{FIGU1}).

Even though Eqs. (\ref{INT3a}) and (\ref{INT3b}) are central to the analytic estimates,
an accurate assessment the cosmic graviton spectrum can be obtained 
in terms of the transfer function of the energy density\footnote{It is customary to introduce 
the transfer function of the power spectrum and the transfer function of the energy density. 
Although the two concepts are complementary, the latter turns out to be more useful 
 than the former when dealing with the cosmic graviton spectrum (see, in particular, the last paper of 
Ref. \cite{foura} for a complete discussion of the problem).}. Across equality the transfer function is
\begin{equation}
T_{eq}(\nu,\nu_{eq}) = \sqrt{1 + c_{eq}\biggl(\frac{\nu_{\mathrm{eq}}}{\nu}\biggr) + b_{eq}\biggl(\frac{\nu_{\mathrm{eq}}}{\nu}\biggr)^2}.
\label{Teq}
\end{equation}
To transfer the spectral energy density inside the Hubble radius Eqs. (\ref{EQ6}) and (\ref{INT1}) are integrated numerically across equality
and this procedure fixes the numerical coefficients  $c_{eq}= 0.5238$ and $b_{eq}=0.3537$ \cite{foura} and the typical 
frequency of the transition:
\begin{equation}
\nu_{eq}  =  1.362 \times 10^{-17} \biggl(\frac{h_{0}^2 \Omega_{\mathrm{M}0}}{0.1411}\biggr) \biggl(\frac{h_{0}^2 \Omega_{\mathrm{R}0}}{4.15 \times 10^{-5}}\biggr)^{-1/2}\,\, \mathrm{Hz}.
\label{nueq}
\end{equation}
The same procedure leading to Eqs. (\ref{Teq}) and (\ref{nueq}) gives the transfer function across the intermediate frequency 
$\nu_{*}$ already introduced in Eq. (\ref{ONEa}):
\begin{equation}
T_{*}(\nu, \nu_{*}) = \biggl[1 + c_{*}\biggl(\frac{\nu}{\nu_{*}}\biggr)^{2\epsilon +  n_{T}} + b_{*}\biggl(\frac{\nu}{\nu_{*}}\biggr)^{4 \epsilon + 2 n_{T}}\biggr]^{-1/2},
\label{Tstar}
\end{equation}
 while $c_{eq}$ and $b_{eq}$ can be accurately assessed, $c_{*}$ and $b_{*}$ depend 
on the parametrization of the refractive index but are of order $1$. Finally the transfer 
function across $\nu_{s}$ determines the high-frequency 
branch of the spectrum 
\begin{eqnarray}
T_{s}(\nu,\nu_{s}) &=& \sqrt{ 1 + c_{s}  \biggl(\frac{\nu}{\nu_{s}}\biggr)^{p(w)/2} + b_{s}  \biggl(\frac{\nu}{\nu_{s}}\biggr)^{p(w)}}, \qquad 
p(w) = 2 - \frac{4}{3w +1},
\label{Ts}\\
\nu_{s} &=&  \sigma^{3(w+1)/(3w -1)} \nu_{\mathrm{max}}, \qquad \nu_{spike} = \nu_{\mathrm{max}}/\sigma, \qquad \sigma =
\biggl(\frac{H_{\mathrm{max}}}{H_{r}} \biggr)^{\frac{1-3w}{6 (w+1)}},
\label{nur}
\end{eqnarray}
where $H_{r}$ denotes the Hubble rate at the onset of the radiation-dominated phase and $w$ is the barotropic index 
of the stiff phase. As in the case of $c_{*}$ and $b_{*}$ also $c_{s}$ and $b_{s}$ change depending on the values of $w$.
In the case $w\to 1$ there are even logarithmic corrections which have been specifically scrutinized in the past (see e.g. \cite{two});
moreover the derivation of $T_{eq}(\nu)$ and $T_{s}(\nu)$, in a different physical situation, has been discussed in detail in the last paper of Ref. \cite{foura}. 

Defining therefore ${\mathcal T}(\nu, \nu_{eq}, \nu_{*}, \nu_{s})$ as the total transfer function, the spectral energy distribution in critical units becomes:
 \begin{eqnarray}
h_{0}^2 \,\Omega_{gw}(\nu,\tau_{0}) &=& {\mathcal N}_{\rho} \,\, r_{T}( \nu_{p})\,\, {\mathcal T}^2(\nu, \nu_{eq}, \nu_{*}, \nu_{s}) \, \,\biggl(\frac{\nu}{\nu_{\mathrm{p}}} \biggr)^{n_{\mathrm{T}}} \, e^{- 2 \,\beta\,\nu/\nu_{\mathrm{max}}}, 
\label{OMA}\\
{\mathcal T}(\nu, \nu_{eq}, \nu_{*}, \nu_{s}) &=& T_{eq}(\nu,\nu_{eq})\, T_{*}(\nu, \nu_{*}) \, T_{s}(\nu, \nu_{s}),
\label{OMB}\\
{\mathcal N}_{\rho} &=& 4.165 \times 10^{-15}\, \biggl(\frac{h_{0}^2 \Omega_{\mathrm{R}0}}{4.15\times 10^{-5}}\biggr) \, \biggl(\frac{{\mathcal A}_{{\mathcal R}}}{2.41\times 10^{-9}}\biggr), 
\label{OMC}
\end{eqnarray}
where $n_{T} = [\alpha ( 3 - 2 \gamma) - 2 \epsilon]/(1 + \alpha - \epsilon)$ and $r_{T}(\nu_{p})$ is the tensor to scalar ratio
evaluated at the pivot frequency $\nu_{p}$:
\begin{eqnarray}
r_{T}(\nu) &=& \epsilon \frac{2^{6 - n_{T}}}{\pi} \Gamma^2\biggl(\frac{3 - n_{T}}{2} \biggr)  e^{q_{T}}\,
\biggl| 1 + \frac{\alpha}{ 1 - \epsilon}\biggr|^{ 2 - n_{T}} \biggl(\frac{\nu}{\nu_{\mathrm{max}}}\biggr)^{n_{T}},
\label{RT}\\
q_{T} &=& ( 3 - 2 \gamma - n_{T})(N_{*}\alpha - \ln{n_{i}}) + n_{T}(N_{t} - N_{*}).
\label{QT}
\end{eqnarray}
In the conventional case $r_{T}(\nu_{p})$ is related to the slow-roll parameter $\epsilon$  and to the tensor spectral index $n_{T}$ via the so-called consistency relations which are however not enforced in the present situation. When the refractive index is not dynamical (i.e. $\alpha \to 0$ and $\gamma \to 0$) 
it is nonetheless true that $n_{T} \to - 2 \epsilon$, as expected. Finally the parameter $\beta={\mathcal O}(1)$ appearing in Eq. (\ref{OMA}) depends upon the width of the transition between the inflationary phase and the subsequent radiation dominated phase;  by using different widths
 we can estimate $0.5 \leq \beta \leq 6.3$ \cite{foura}: the results on slopes of the four branches are not affected by the value of $\beta$ which 
 however controls the rate of exponential suppression after the endpoint frequency.
 
According to Eq. (\ref{Teq}), $T_{eq}(\nu) \to 1$ for $\nu \gg \nu_{eq}$ however the effect of neutrino free-streaming 
introduces a minor source of supplementary suppression in the range $ \nu_{eq} < \nu < \nu_{bbn}$ where $\nu_{bbn}$ denotes the nucleosynthesis frequency (i.e. $ \nu_{bbn} = {\mathcal O}(10^{-11})$ Hz). The neutrino free-streaming produces an effective anisotropic stress leading ultimately to an integro-differential equation\footnote{If the only collisionless species are the neutrinos (which are massless in the concordance paradigm), the amount 
of suppression of $h_{0}^2 \, \Omega_{gw}$  can be parametrized by the function ${\mathcal F}(R_{\nu}) = 1 -0.539 R_{\nu} + 0.134 R_{\nu}^2$ .  This means that we are talking about a figure of the order of ${\mathcal F}^2(0.405)= 0. 645$ (for $N_{\nu} = 3$ and $R_{\nu} = 0.405$).} \cite{neutr}.  This effect is not central to the present discussion but it can be easily 
included; similarly another potential effect is associated 
with the variation of the effective number of relativistic species; in the case of the minimal standard model this would imply that the reduction will be ${\mathcal O}(0.38)$ (see e.g. the last paper of Ref. \cite{two}).  Note finally that if the 
various scales only reenter during radiation we have that $T_{s}(\nu)\to 1$ (or, more formally, $\nu_{s} \to \infty$ for a fixed comoving 
frequency $\nu$).

The parameters of the cosmic graviton spectra illustrated in Fig. \ref{FIGU1} have not been 
randomly guessed but they are consistent with the phenomenological constraints and with 
some basic detectability requirements that will now be elucidated. 
In the low-frequency range the tensor to scalar ratio of Eq. (\ref{RT}) 
is bounded from above not to conflict with the observed temperature and polarization 
anisotropies of the Cosmic Microwave Background; 
we specifically required $r_{T}(\nu_{p}) <0.06$, as it follows 
from a joint analysis of Planck and BICEP2/Keck array data \cite{five}. 
The pulsar timing measurements impose instead a limit at the frequency 
$\nu_{pulsar} \simeq \,10^{-8}\,\mathrm{Hz}$(roughly corresponding to the inverse of the observation 
time along which the pulsars timing has been monitored \cite{six})
 and implying $\Omega_{gw}(\nu_{pulsar},\tau_{0}) < 1.9\times10^{-8}$. 
Finally the big-bang nucleosynthesis sets an indirect constraint  
on the extra-relativistic species (and, among others, on the relic gravitons) at the time when light nuclei 
have been formed \cite{seven}. This constraint is often expressed in terms of $\Delta N_{\nu}$ 
representing the contribution of supplementary (massless) neutrino 
species (see e.g. \cite{sevena}) but the extra-relativistic species do not need to be fermionic. 
If, as in our case, the additional species are relic gravitons we will have to demand that:
\begin{equation}
h_{0}^2  \int_{\nu_{bbn}}^{\nu_{\mathrm{max}}}
  \Omega_{gw}(\nu,\tau_{0}) d\ln{\nu} = 5.61 \times 10^{-6} \Delta N_{\nu} 
  \biggl(\frac{h_{0}^2 \Omega_{\gamma0}}{2.47 \times 10^{-5}}\biggr).
\label{BBN1}
\end{equation}
The bounds on $\Delta N_{\nu}$ range from $\Delta N_{\nu} \leq 0.2$ 
to $\Delta N_{\nu} \leq 1$ so that the right hand side of Eq. (\ref{BBN1}) 
turns out to be between $10^{-6}$ and $10^{-5}$. 
\begin{figure}[!ht]
\centering
\includegraphics[height=7.2cm]{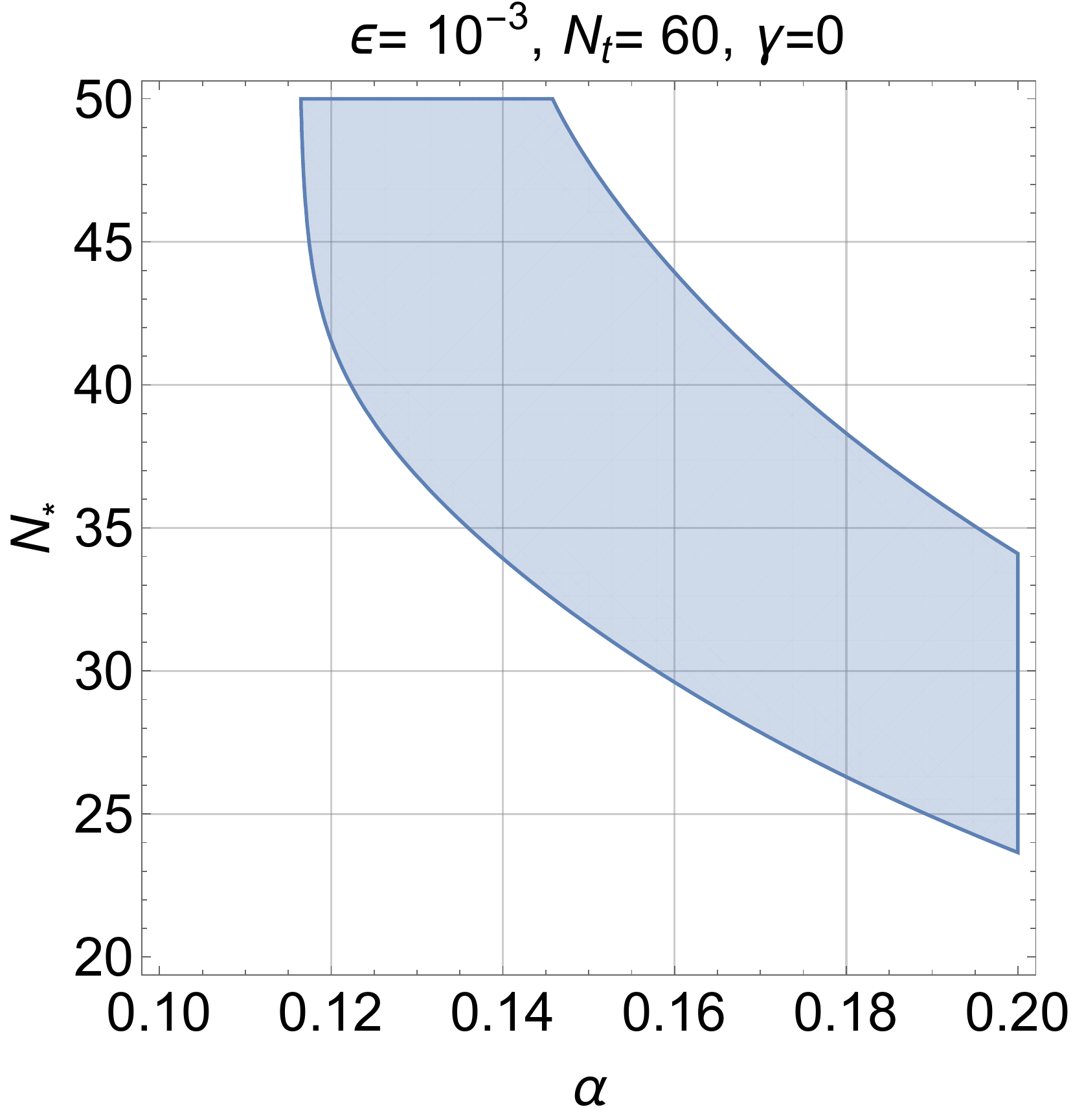}
\includegraphics[height=7.2cm]{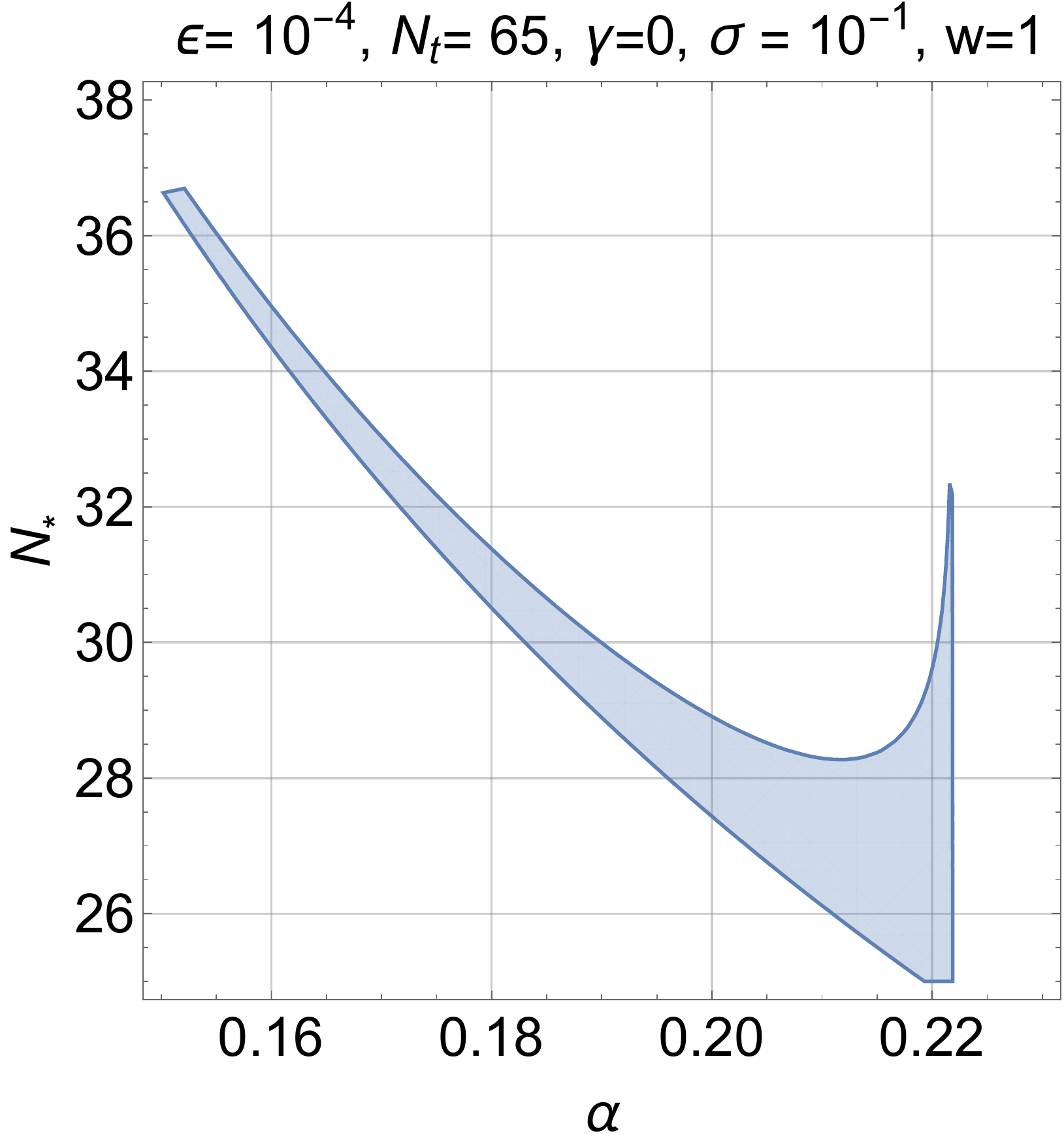}
\caption[a]{In the shaded regions the phenomenological constraints are all satisfied and the corresponding spectral energy density 
leads to a potentially detectable signal both in the mHz and in the kHz bands. }
\label{FIGU2}      
\end{figure}
The shaded areas of Fig. \ref{FIGU2} illustrate the regions where not only the phenomenological constraints 
are concurrently satisfied but the spectral energy density is also potentially detectable both in the mHz band 
(i.e. $ 0.1 \, \mathrm{mHz}<\nu_{mHz} < \mathrm{Hz}$) and in the audio band (i.e. $\mathrm{Hz} 
< \nu_{audio} < 10\, \mathrm{kHz}$). 
In particular we required $h_{0}^2 \Omega_{gw}(\nu_{audio},\tau_{0}) \geq 10^{-10}$ 
hoping that (in a not too distant future) the Ligo/Virgo detectors in their advanced configurations 
will reach comparable sensitivities \cite{eight}. With a similar logic we are led to require
$h_{0}^2 \Omega_{gw}(\mathrm{mHz}, \tau_{0}) \geq 10^{-12}$ 
in the mHz band always assuming that comparable sensitivities will be reached by space-borne 
interferometers \cite{nine} which are, at the moment, only proposed and not yet operational.
In the left plot of Fig. \ref{FIGU2} we illustrate the case where all the modes reenter during 
radiation while in the right plot we consider the presence of a stiff phase preceding the ordinary
 radiation epoch. In case the onset of the radiation-dominated phase is delayed by the presence of a stiff phase; 
a spike appears in the GHz region and the signal is comparatively more constrained: this is the reason why 
the area of the left plot is larger than the area of the right plot.  This kind of signal might however be interesting 
for electromagnetic detectors of gravitational radiation which have been proposed and partially developed
through the past decade \cite{ten}. As anticipated the parameters of the two plots of Figs. \ref{FIGU1} have been drawn, respectively,
from the shaded areas of the two plots illustrated in Fig. \ref{FIGU2}. 
 
The inflationary scenarios based on a quasi-de Sitter stage of expansion 
suggest that the spectral energy distribution of the cosmic gravitons should always decrease in frequency 
and hence remain below $10^{-17}\, \rho_{crit}$ both in the mHz and in the kHz bands. 
If the refractive index of the tensor modes is dynamical $\Omega_{gw}(\nu,\tau_{0})$ develops an increasing branch at intermediate frequencies
while it flattens out above the $\mu$Hz region with an approximate amplitude that can exceed 
the conventional signal even by nine orders of magnitude. In this case the quasi-flat plateau present in the conventional situation 
gets larger and it is pushed at higher frequencies. All in all the increase of the spectral energy density does not conflict with the limits 
applicable to the cosmic graviton backgrounds and leads to potential signals both in the audio and in the mHz windows.
If the onset of the radiation epoch is delayed by a post-inflationary phase with equation of state stiffer than 
radiation, the spectral energy density inherits a further spike in the GHz region. 
Potentially detectable signals can then be expected both for terrestrial interferometers and for space-borne 
detectors provided the refractive index is dynamical at least for a limited amount of time during an otherwise conventional 
quasi-de Sitter stage of expansion. 

It is a pleasure to acknowledge useful discussions with F. Fidecaro. 
The author wishes also to thank T. Basaglia, A. Gentil-Beccot and 
S. Rohr of the CERN Scientific Information Service for their kind assistance.

\end{document}